\documentclass[12pt]{article}

\usepackage{lmodern,color,soul}

\usepackage{blindtext}

\usepackage{scicite}

\usepackage{times}
\usepackage{siunitx}
\usepackage{graphicx}
\usepackage{float}
\usepackage{caption}
\usepackage{subcaption}

\title{IRIS: Integrated Retinal Functionality in Image Sensors}

\date{}

\begin{document} 


\baselineskip24pt


\maketitle 
\paragraph*{FRONT MATTER}
\section*{IRIS: Integrated Retinal Functionality in Image Sensors}

\section*{Authors}
Zihan Yin$^{1\dagger}$, Md Abdullah-Al Kaiser$^{1\dagger}$, Lamine Ousmane Camara$^{2}$, Mark Camarena$^{1\dagger}$, Maryam Parsa$^{3}$, Ajey Jacob$^{1}$, Gregory Schwartz$^{2}$, Akhilesh Jaiswal$^{1\ast}$\\
\normalsize{$^{\dagger}$:these authors contributed equally to this work}\\

\section*{Affiliations}

 

\normalsize{$^{1}$ Information Sciences Institute, University of Southern California}\\
\normalsize{$^{2}$ Department of Ophthalmology, Northwestern University}\\
\normalsize{$^{3}$ Electrical and Computer Engineering, George Mason University}\\
\normalsize{$^{\ast}$Corresponding author: akjaiswal@isi.edu}\\

\section*{Abstract}
    
Neuromorphic image sensors draw inspiration from the biological retina to implement visual computations in electronic hardware. Gain control in phototransduction and temporal differentiation at the first retinal synapse inspired the first generation of neuromorphic sensors, but processing in downstream retinal circuits, much of which has been discovered in the past decade, has not been implemented in image sensor technology. We present a technology-circuit co-design solution that implements two motion computations occurring at the output of the retina that could have wide applications for vision based decision making in dynamic environments. Our simulations on Globalfoundries 22nm technology node show that, by taking advantage of the recent advances in semiconductor chip stacking technology, the proposed retina-inspired circuits can be fabricated on image sensing platforms in existing semiconductor foundries. Integrated Retinal Functionality in Image Sensors (IRIS) technology could drive advances in machine vision applications that demand robust, high-speed, energy-efficient and low-bandwidth real-time decision making.  




\section*{Introduction}
Animal eyes are extremely diverse and specialized for the environment and behavioral niche of each species \cite{land2005optical}. Specialization is particularly robust in the retina, a part of the central nervous system containing parallel circuits for representing different visual features. In contrast, the engineered `eyes,' \textit{i.e.} image sensor technology, used in machine vision, are highly stereotyped. Even though cameras can have different optics on the front end, the image sensor chip, which represents the electronic analogue of the biological retina, is essentially a two-dimensional array of pixels with each transmitting a luminance signal at a fixed frame rate \cite{1175960}. A motivating hypothesis for the present work is that the efficiency and performance of machine vision can be improved using specialized image sensors that replicate some of the feature-selective computations in biological retinas. 

Rod and cone photoreceptors form the input layer of the vertebrate retina where they transduce light into an analog voltage signal that is then transmitted via \textit{bipolar cells} to the inner retina. Signals diverge at this first synapse from each cone photoreceptor onto approximately 15 different bipolar cell types \cite{eggers2011multiple}. Functional divergence and the sophistication of visual processing then increase dramatically in the inner retina where more than 60 \textit{amacrine cell} types \cite{sanes_60_amacrines} shape the signals to implement various computations. Finally, signals from bipolar and amacrine cells are collected by over 40 types of \textit{retinal ganglion cells }(RGCs), the output cells of the retina whose axons form the optic nerve \cite{rgc_types}. 
RGCs transmit spike trains that carry information about specific visual features like object movement, direction, orientation, and color contrast \cite{sernagor2001development}. Each RGC type provides a full representation of visual space. Thus, while the input layer of the retina is analogous to an analog pixel array (albeit one with built-in filtering and gain control), once the photoreceptor signals have been processed by the dozens of cell types comprising retinal circuits, the output representation is very different, representing specific visual features. Binary RGC spike trains convey information about more than 40 different visual features to the brain, and each point in visual space is represented in parallel in all of the feature-selective RGC outputs.    

Efforts to bring biologically-inspired functionality to electronic image sensors date back to at least the 1980s \cite{mead1988silicon} followed with the advent of neuromorphic sensors (reviewed in \cite{liu2010neuromorphic},\cite{zhu2020technologies}). Two related aspects of visual computation, which were already well characterized in retinal neurobiology by the late 1980s, have dominated the field of neuromorphic vision sensors. The first idea was to mimic luminance adaptation, the computation used by the retina to adjust the dynamic range of its biological components to that of the visual scene. Humans use vision over 10 orders of magnitude of luminance \cite{rodieck1998first} and even single natural images vary in brightness by more than a factor of $10^5$ \cite{FRAZOR20061585}. These high dynamic ranges are poorly represented by linear photodiodes and digitization to 8 or even 12 bits. High dynamic range (HDR) cameras use multiple exposures to reconstruct an image, trading bit depth for frame rate \cite{schanz2000high}, while logarithmic detectors use range compression to avoid saturation \cite{log_sensor}. The second aspect of retinal computation to take hold in neuromorphic image sensors is change detection, the propensity of retinal neurons to adapt to the mean luminance over time and only transmit information about its change. Event-based cameras, or Dynamic Vision Sensors (DVS), implement temporal differentiation at each pixel and asynchronously transmit binary `spike' events when the luminance change exceeds a threshold.  The asynchronous transmission of DVS cameras has critical advantages for high speed operation, since it is not limited by frame rate, and for efficiency, since pixels that do not change do not transmit data (reviewed in \cite{ETIENNECUMMINGS199619, Liao_2021}).

\begin{figure}[!t]
\centering
\includegraphics[width =1.05\linewidth]{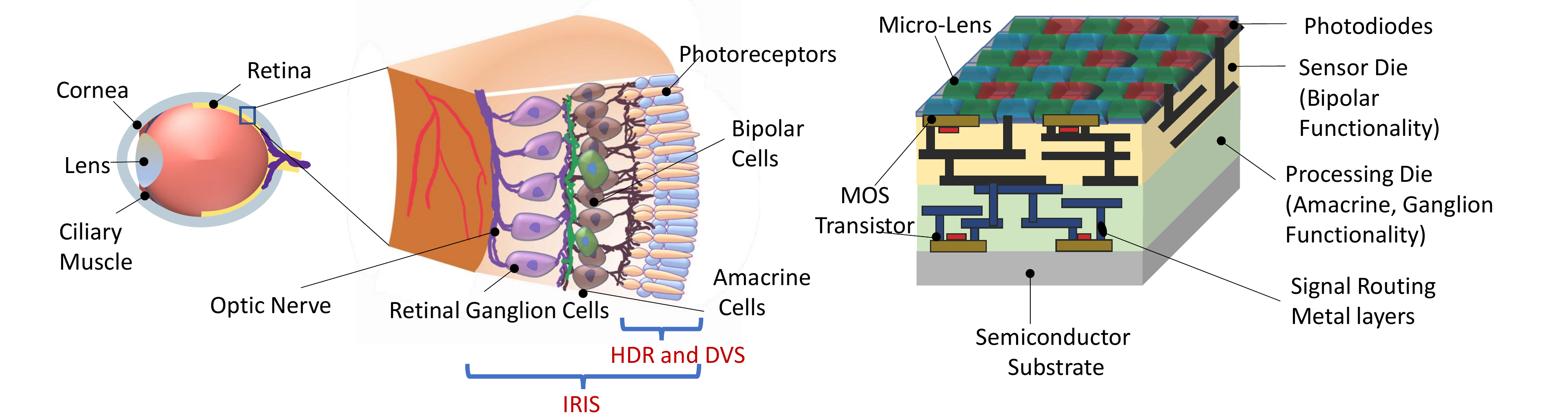}
\caption{(Left) Representational view of biological retina. (Right) Proposed IRIS camera implementing retinal computations on a back-side illuminated active pixel sensor \cite{sony_BI_3D} camera with Bayer pattern color filters \cite{catrysse2005roadmap}. Two retina inspired functionality found in existing works are HDR (High Dynamic Range) and DVS (Dynamic Vision Sensor), which are confined to phototransduction and first retinal synapse. In contrast, the proposed IRIS camera aims to implement computations that occur in the circuits of inner retina, mimicking the \textit{feature-spikes} of Retinal Ganglion Cells (RGC).}
\label{overview}
\vspace{-1mm}
\end{figure}

In this work, we present a new class of neuromorphic sensors called \textit{Integrated Retinal Functionality in Image Sensors (IRIS)}. By leveraging recent advances in the understanding of inner retinal circuits, IRIS technology goes beyond luminance adaptation and change detection, features mostly confined to phototransduction and the first retinal synapse, to implement computations that occur in the circuits of the inner retina, mimicking the feature-selective spike trains of RGCs. Here we present IRIS circuits implementing two retinal motion computations: \textit{Object Motion Sensitivity} (OMS) and \textit{Looming Detection} (LD). Note, the aim of the present work is not to implement the detailed electro-chemical dynamics of retinal cell types, rather to functionally mimic the computational behavior of retinal circuits on state-of-the-art image sensing platforms. 

OMS is a computation that enables the visual system to discriminate motion of objects in the world (object motion) from motion due to one's own eye, head, and body movements (self motion) (reviewed in \cite{baccus2008retinal}). A subset of RGCs respond to either local motion in the receptive field `center' or differential motion of the receptive field `center' and `surround' regions but remains silent for global motion \cite{yu2019moving}. OMS RGCs are thought to be important in detecting movements of predators and prey amidst a background of self motion \cite{schwartz2021retinal}. For machine vision applications a fast sensor with built-in OMS could detect moving objects even if the camera itself was moving, for example, on an autonomous vehicle. 

LD is a computation that likely evolved to warn animals of approaching threats, especially those from overhead (reviewed in \cite{temizer2015visual}). Loom-sensitive RGCs respond selectively to expanding (approaching) dark objects with much weaker responses to translational motion across the visual field \cite{munch_looming}. Experiments in flies \cite{Card:2012tf}, zebrafish \cite{temizer2015visual}, frogs \cite{Ishikane2005}, and mice \cite{yilmaz_meister_looming,Wang_2021_looming} have established a causal role for LD RGCs in eliciting stereotyped escape responses. In machine vision, an LD-equipped sensor could be used on an autonomous vehicle to avoid collisions by enabling fast detection of approaching objects. 

We show OMS and LD circuits built on standard Complementary Metal Oxide Semiconductor (CMOS) technology based on active pixel sensors (APS) as well as DVS pixels. We exploit advances in semiconductor chip stacking technology and highly scaled, dense CMOS transistors, to embed retina-inspired circuits in a hierarchical manner analogous to the processing layers of the biological retina. Our simulations demonstrate the prevalence of OMS and LD triggering stimuli in natural scenes from moving vehicles, and they show circuit designs that implement both the OMS and LD computations and are compatible with existing image sensor fabrication technology. This work forms the necessary foundation to build IRIS-equipped cameras for machine vision.

\section*{Results}

\subsection*{Algorithmic Implementation of Retinal Computations} 
Feature selective circuits in the vertebrate retina, like OMS and LD, are built from 5 classes of neurons (Fig. \ref{overview}). Photoreceptors form the input layer (like the pixels in a camera) and retinal ganglion cells (RGCs) represent the output. The computations that transform the pixel-like representation of the photoreceptors to the feature selective representation of RGCs are carried out by the 3 interneuron classes: horizontal cells, bipolar cells, and amacrine cells. Horizontal cells are mainly involved in lateral inhibition and color processing, but they do not play a major role in OMS and LD circuits \cite{schwartz2021retinal}. Thus, we designed the components of IRIS circuits to match the functionality of bipolar, amacrine and ganglion cells in these computations.

\begin{figure}[!t]
\centering
\includegraphics[width =\linewidth]{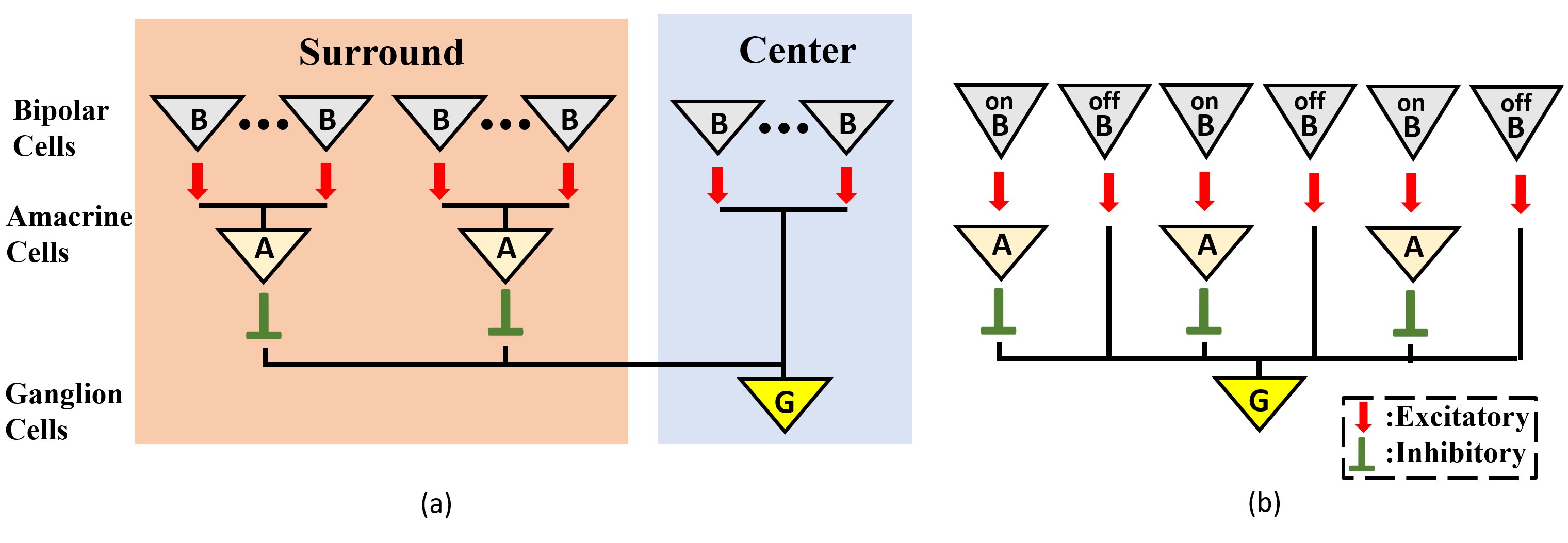}
\caption{(Left) Retinal Object Motion Sensitive Circuit. (Right) Retinal Looming Detection Circuit.}
\label{biocircuit}
\vspace{-1mm}
\end{figure}

Both computations begin with bipolar cells that act like differentiators; they adapt (rapidly) to steady illumination and signal only changes in luminance. In the biological retina, separate bipolar cells carry signals for positive (ON) and negative (OFF) changes in illumination. The OMS retinal circuit (Fig. \ref{biocircuit}(a)) combines this functionality at the level of ON-OFF bipolar-like units, while the LD circuit Fig. \ref{biocircuit}(b) has separate ON and OFF bipolar sub-circuits.

\begin{figure}[!h]
\centering
\includegraphics[width =\linewidth]{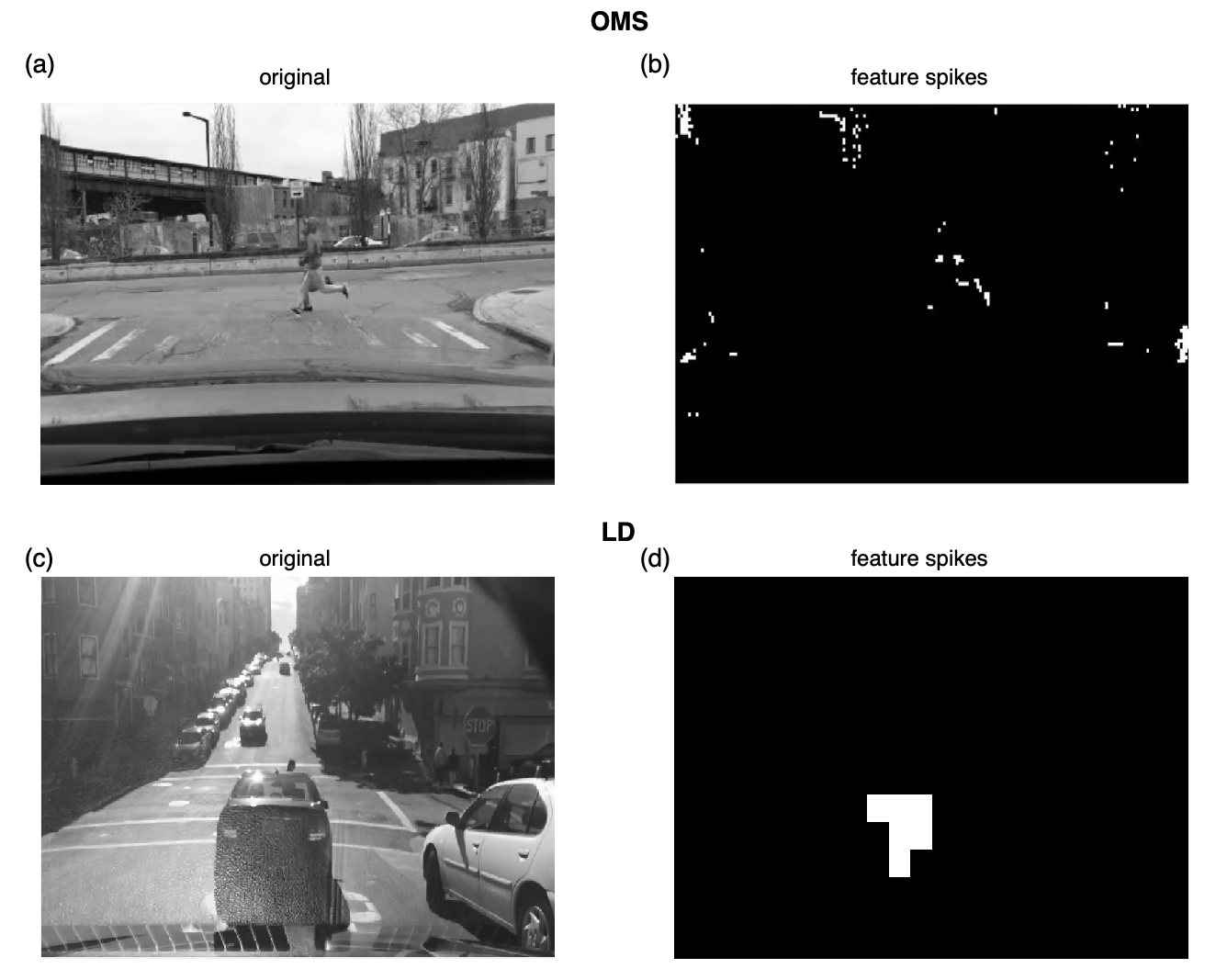}
\caption{Example frames showing the software implementation of the OMS and LD algorithms. Images in (a) and (c) are frames from the Berkeley DeepDrive database \cite{bdd100k}. Frames in (b) and (d) show the corresponding sparse representation at the output of the OMS and LD circuits, respectively. White pixels indicate spike events. Spikes can be seen in frame (b) indicating identification of differential motion (runners in front of a moving car) in accordance with the expected OMS behavior. Similarly, frame (d) shows spikes indicating a looming (approaching) car in the receptive field.}
\label{ExampleFrames}
\vspace{-1mm}
\end{figure}

Amacrine cells are the most diverse class of neurons in the retina, comprising more than 60 cell types \cite{sanes_60_amacrines}. While many of the cellular and synaptic details of amacrine cells remain incompletely understood, their algorithmic role in the OMS and LD circuits has been well characterized \cite{baccus2008retinal, zhang_2012_looming}, reviewed in (\cite{schwartz2021retinal}). In the OMS circuit, amacrine cells collect the bipolar cells' contrast responses from a wide spatial extent, the receptive field `surround', and relay this summed signal with an opposite sign to the output of bipolar cells from the receptive field `center', implementing a spatial filter with a subtraction operation, thereby detecting differential motion between the `center' and the `surround' regions. In the LD circuit, amacrine cells also invert the sign of signals from bipolar cells, but on a smaller spatial scale. OFF signals from the leading edge of a dark moving object are relayed directly by OFF bipolar cells to RGCs, while ON signals from the trailing edge of the object are relayed with opposite sign to the RGC via intermediary amacrine cells. Thus, moving objects with both leading and trailing edges elicit opposing responses that cancel at the level of the RGC, while expanding dark objects that have only leading edges, elicit an RGC response. A more detailed description of the functioning of OMS and LD circuits can be found in \cite{schwartz2021retinal} and \cite{eye_smarter_review}.

Before designing the electronics for IRIS sensor hardware, we confirmed our algorithmic understanding of the OMS and LD computations by implementing them in software and testing them on dashboard video segments from the Berkeley DeepDrive database \cite{chen2015deepdriving}. The OMS algorithm elicited simulated RGC spikes for the expected features of the videos, like runners crossing the street in front of the moving car (Fig. \ref{ExampleFrames}(a,b)). Likewise, the LD circuit signaled the approach of negative contrast (dark) objects (Fig.  \ref{ExampleFrames} (c,d)). Based on these results, we sought to design hardware IRIS circuits using mixed signal design that implemented the same OMS and LD computations as our software.

\section*{Embedding OMS Functionality in Image Sensors} 

\begin{figure}[!t]
\centering
\subfloat[]{\includegraphics[width=0.9\linewidth]{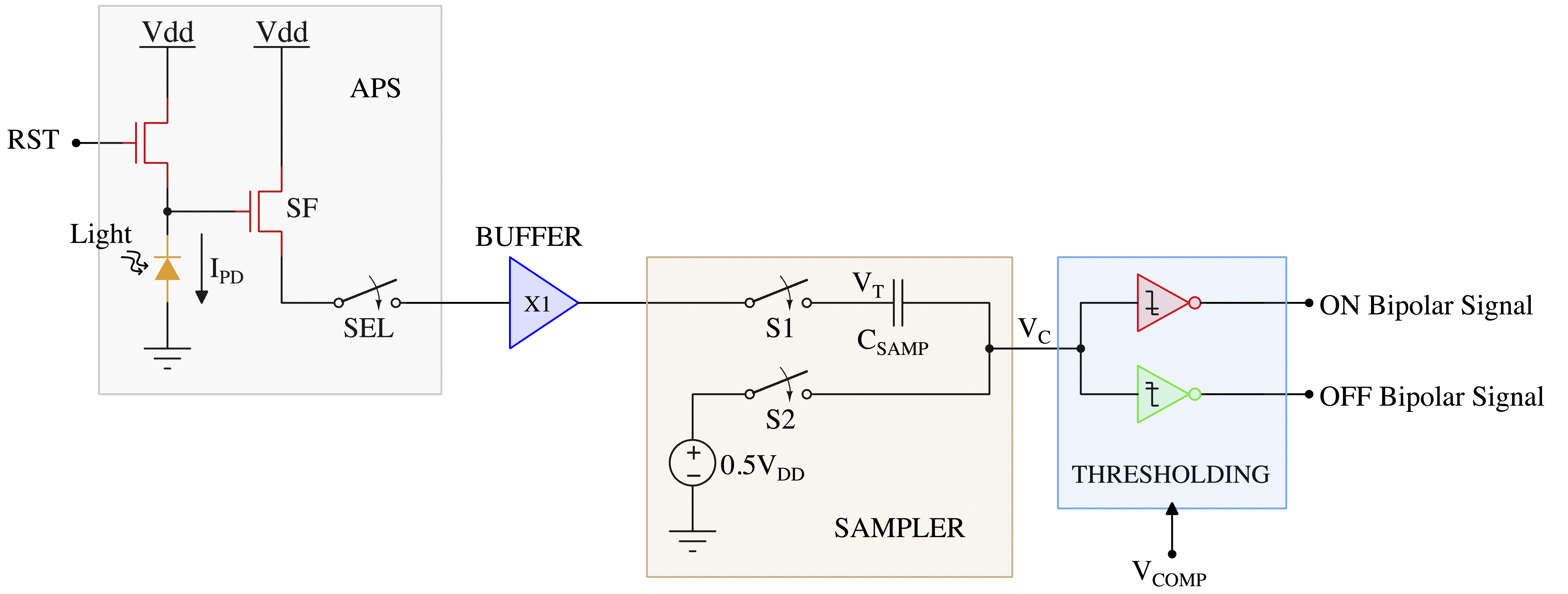}}
\newline
\subfloat[]{\includegraphics[width=0.6\linewidth]{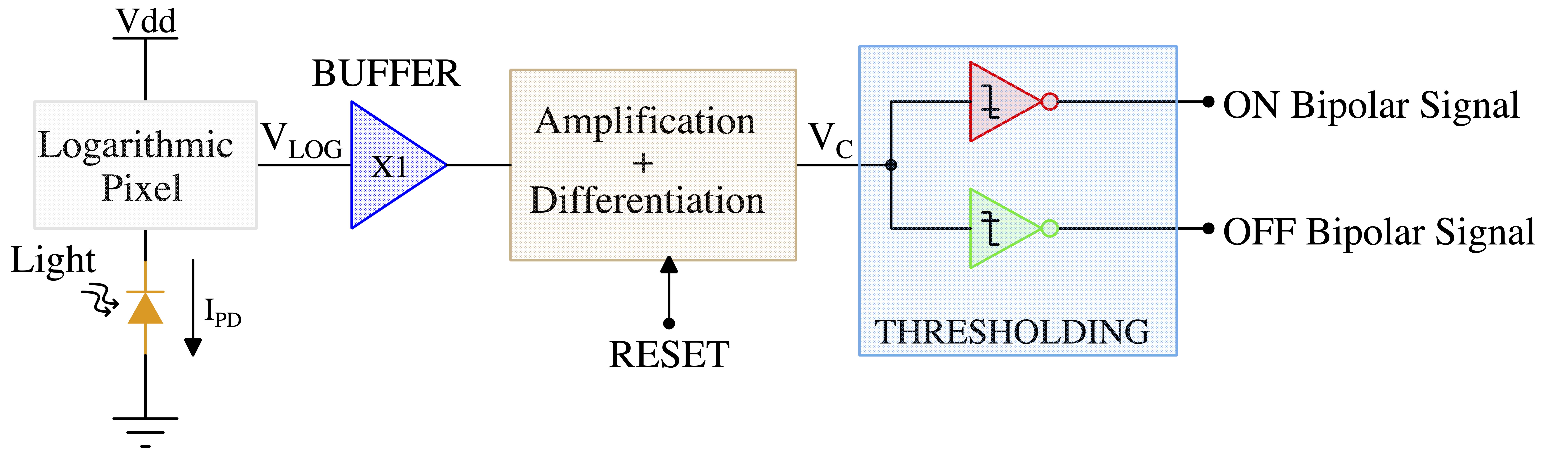}}
\subfloat[]{\includegraphics[width=0.4\linewidth]{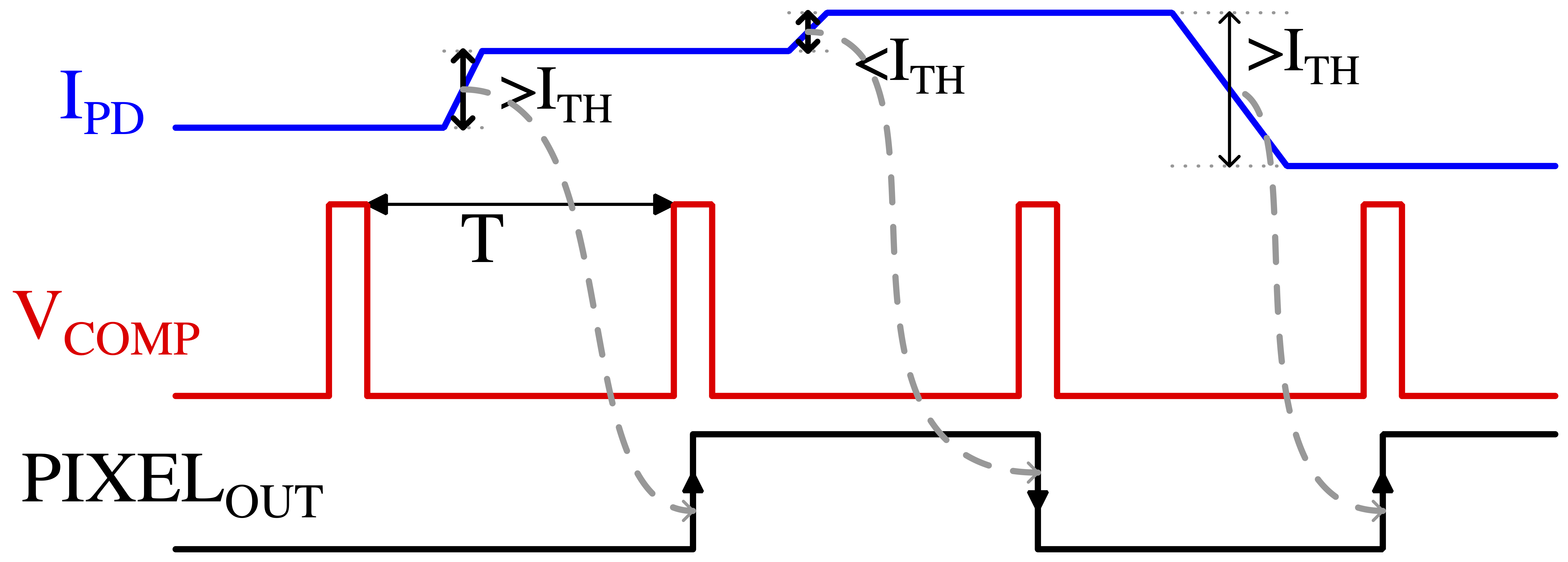}}
\caption{(a) CMOS implementation of the APS Pixel circuit diagram and (b)  DVS Pixel circuit diagram and (c) The timing waveform of the retinal bipolar functionality using APS and DVS Pixel.}
\label{pixel_ckt}
\end{figure}

As described above, the OMS computation in the retina starts by detection of change in temporal contrast of input light by the bipolar cells. In other words, for OMS behavior, functionally, the bipolar cells generate an electrical signal for a change in light intensity above a certain threshold. Fig. \ref{pixel_ckt}(a)-(b) show solid-state circuits that can mimic the bipolar cell's contrast sensitive behavior using conventional CMOS Active Pixel Sensor (APS) \cite{chi2007cmos} and Dynamic Vision Sensor (DVS) \cite{samsung_isscc17}, respectively. Note, APS pixels are of specific importance since they form the backbone of state-of-the-art camera technology \cite{samsung_32MP} and wide class of computer vision applications \cite{voulodimos2018deep}.

For APS-based implementation the focal plane array is formed by a 2 dimensional array of APS pixels with additional circuit to enable temporal light contrast-change detection. The array of such contrast-change sensitive APS pixels sample the input light intensity for each frame, in parallel, and compare it to the light intensity of the next frame. If the light intensity sensed by each APS pixel increases (decreases), the contrast sensitive APS pixels would generate an ON (OFF) \textit{bipolar-signal}.
Consider the APS-based contrast-change detection circuit of Fig. \ref{pixel_ckt}(a). For the APS pixels, the output voltage of the well-known 3 transistor pixel circuit is inversely linear proportional to the incident light intensity \cite{kleinfelder200110000}. A source follower buffer (\si{X1}) isolates the sensitive pixel node from the noisy switches (S1) and (S2) in the SAMPLER block. The SAMPLER block samples the pixel output, in parallel for each frame, and performs analog subtraction operation between two consecutive samples (or frames) for each pixel, simultaneously. The subtraction operation starts by sampling the buffered 3T APS pixel voltage of the first frame on the top-plate and a constant 0.5\si{V_{DD}} on the bottom plate of the sampling capacitor (\si{C_{SAMP}}). In the next frame, the bottom plate of the sampling capacitor is left floating, whereas, the top-plate samples the consecutive frame's pixel voltage. As a result, the floating bottom plate of the capacitor (node \si{V_C}) follows the top plate of the capacitor and stores the difference voltage of the two consecutive frames offset by a constant voltage of 0.5\si{V_{DD}}. Finally, the difference voltage (corresponding to the intensity or contrast change for a given pixel between two consecutive frames) on the bottom-plate of the sampling capacitor is compared to a threshold using the THRESHOLDING circuit (implemented using two transistor static inverter based comparators \cite{samsung_isscc17}). The THRESHOLDING circuit generates a spike through the ON (OFF) channel if the light intensity has increased (decreased) between two consecutive frames. Note, the array of contrast-sensitive APS pixels operate \textit{synchronously} (when the \si{V_{COMP}} is HIGH) generating a \textit{bipolar-signal} for changes in light intensity between two consecutive frames.

In the DVS-based contrast-sensitive pixel circuit (Fig. \ref{pixel_ckt}(b)), a logarithmic photoreceptor transduces the incident light into a logarithmic output voltage \cite{pardo2015selective}. Similar to the APS-based circuit, the source follower buffer (\si{X1}) isolates the sensitive pixel node and the following difference amplifier. The difference amplifier is implemented as a capacitive feedback amplifier that can calculate the gradient of voltage corresponding to incident light intensity change in an \textit{asynchronous } manner \cite{lichtsteiner2006128}. Finally, the output voltage from the difference amplifier is compared in the THRESHOLDING circuit that is similar to the APS-based circuit and generates the ON/OFF \textit{bipolar-signal}. 

Fig. \ref{pixel_ckt}(c) presents a representative timing waveform of the APS pixel-based \textit{bipolar-signal }generation circuit. \si{I_{PD}} represents the photodetector current corresponding to the incident light, and \si{PIXEL_{OUT}} refers to the ON/OFF \textit{bipolar-signal} from the pixel circuit. \si{V_{COMP}} enables the comparison between two consecutive frames with a period of T (depends on the video frame rate). It can be observed from the figure that when the photodetector current (corresponding light intensity) difference in both directions is higher (lower) than the threshold (\si{I_{TH}}), \si{PIXEL_{OUT}} generates a high (low) signal output and that is updated according to the frame rate. Only the waveform of the APS pixel-based circuit has been shown  as most of today's commercial cameras are using the APS pixel. However, the timing waveform of the DVS pixel-based circuit is similar to the APS pixel-based circuit, except that DVS pixels generate asynchronous spikes and is not based on timing of signal \si{V_{COMP}}. 

\begin{figure}[!t]
\centering
\subfloat[]{\includegraphics[width=0.8\linewidth]{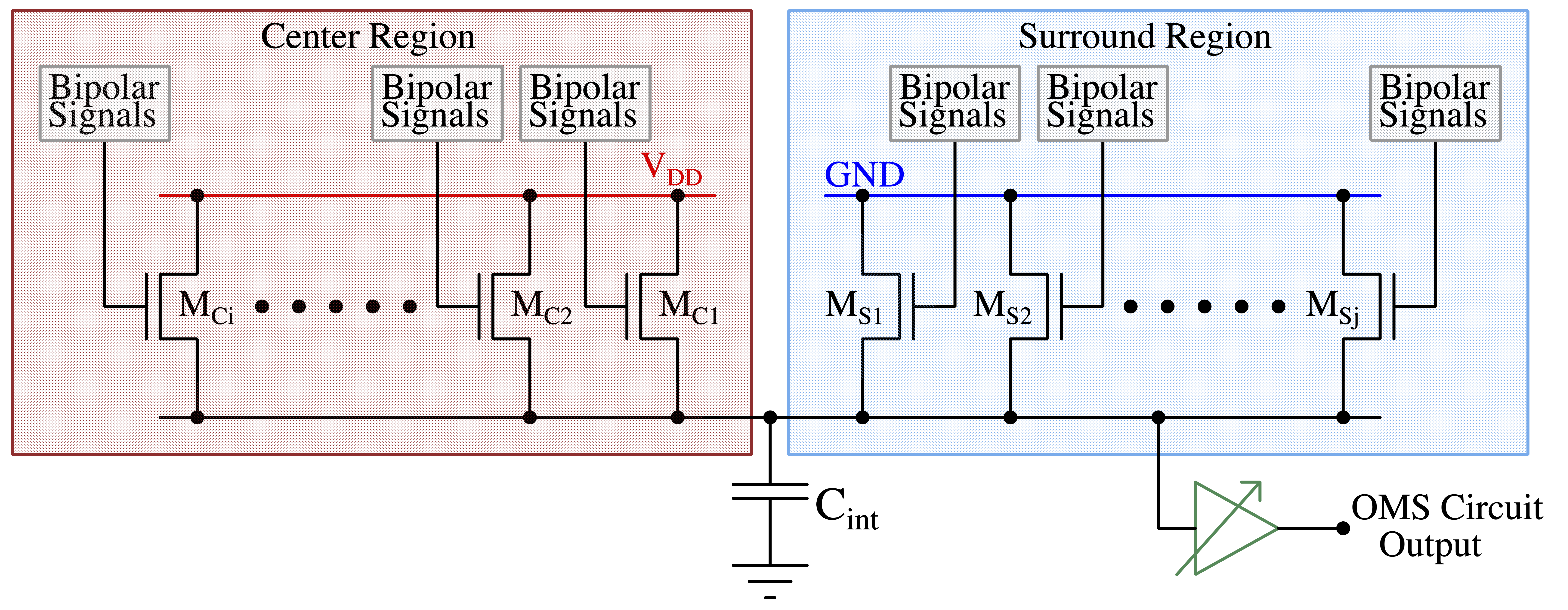}}
\newline
\vspace{-0.1in}
\subfloat[]{\includegraphics[width=0.45\linewidth]{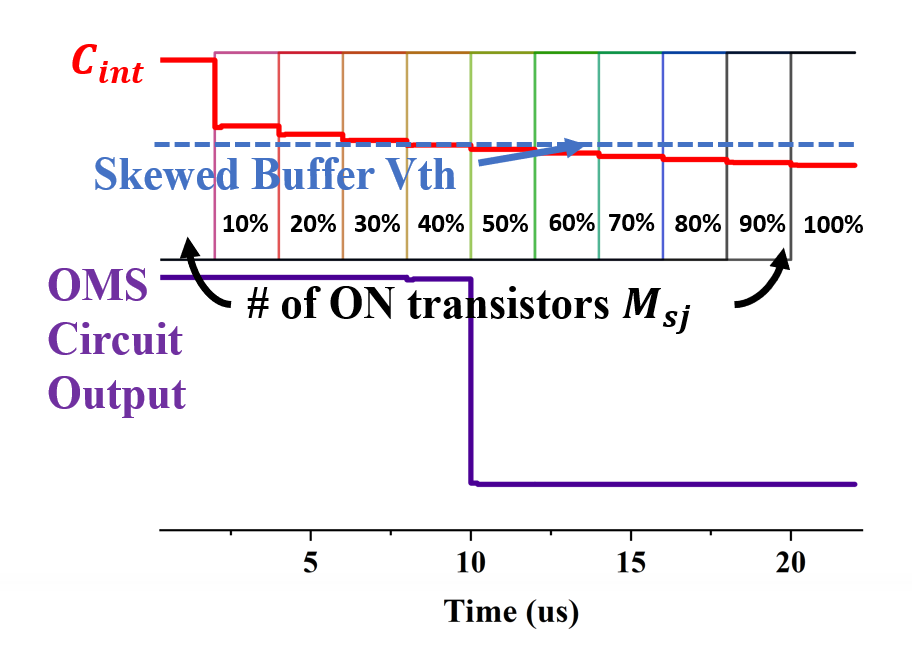}}
\hspace{0.05\textwidth}
\subfloat[]{\includegraphics[width=0.45\linewidth]{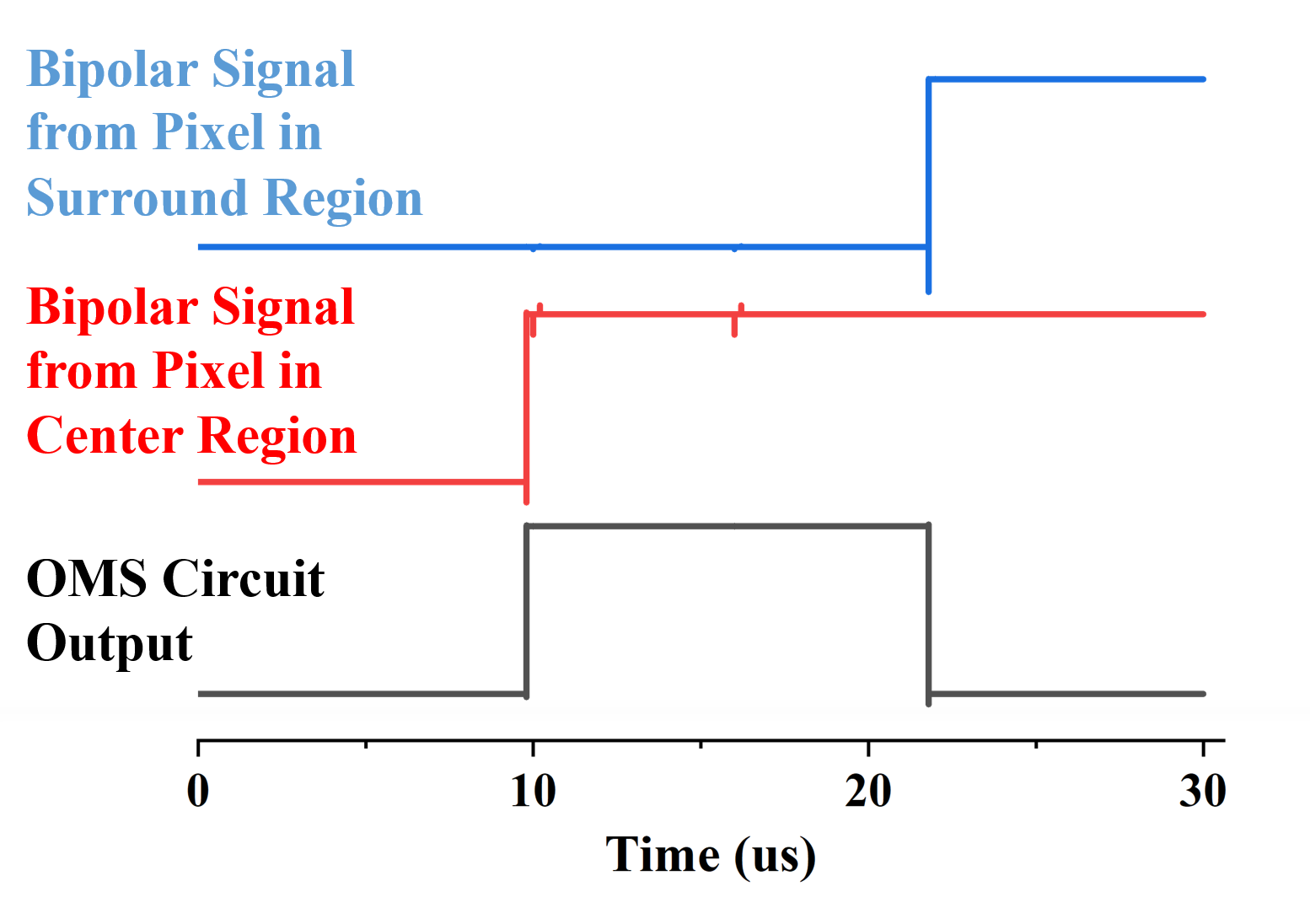}}
\caption{(a) CMOS implementation of the OMS circuit. (b) Voltage on node \si{C_{int}} as a function of number of ON `surround' transistors while keeping all `center' transistors ON. \si{V_{th}} represents the trip point of the buffer whose output represents OMS feature-spike.  (c) Typical timing waveform with one `center' and one `surround' pixel.}
\label{oms}
\end{figure}

The \textit{bipolar-signals} generated from each pixel (either APS-based or DVS-based) are further processed by the circuit shown in Fig. \ref{oms}(a), which implements the functionality of amacrine and ganglion cells for generation of \textit{OMS-feature-spikes}. The circuit of Fig. \ref{oms}(a) consists of two groups of transistors, those belonging to the `center' region (transistors \si{M_{Ci}}s) and those belonging to the `surround' region (transistors \si{M_{Si}}s) in the receptive field. The gate of the `center' (`surround') region transistors \si{M_{Ci}}s (\si{M_{Si}}s) are driven by \textit{bipolar-signals} generated from pixels belonging to the `center' (`surround') region. Further, the upper terminal (drain) of the `center' region transistors are connected to supply voltage \si{V_{DD}}, while the upper terminal (source) of the `surround' region transistors are connected to ground. This ensures when a particular `center' transistor \si{M_{Ci}} receives a \textit{bipolar-signal}, it is switched ON and \textit{integrates} charges on capacitor \si{C_{int}}. Higher the number of \textit{bipolar-signals} generated from the center region, higher would be the resultant voltage on the capacitor \si{C_{int}}. Conversely, when a specific transistor \si{M_{Si}} receives a \textit{bipolar-signal} from `surround' region, it turns ON and attempts to drain the charge stored on the capacitor \si{C_{int}} through the ground terminal. Higher the number of \textit{bipolar-signals} received by the `surround' transistors, lower the resultant voltage on the capacitor \si{C_{int}}. Essentially, the group of transistors \si{M_{Ci}}s and \si{M_{Si}}s form a voltage divider that dictates the resultant voltage on \si{C_{int}}. The voltage on \si{C_{int}} drives a high-skewed CMOS buffer, which generates a spike if the voltage on \si{C_{int}} exceeds the threshold voltage (or the trip-point) of the CMOS buffer.

In summary, when the pixels in the `center' region generate \textit{bipolar-signals}, and at the same time, pixels in the `surround' region also generate \textit{bipolar-signals}, it indicates that the receptive field comprising of the `center' and the `surround' region is experiencing global or background motion without any object motion. In such a case, the voltage accumulated on the capacitor \si{C_{int}} from  transistors \si{M_{Ci}}s in the `center' region are offset by the discharge effect of transistors \si{M_{Si}}s in the `surround' region and the buffer output remains low. However, if the `center' transistor \si{M_{Ci}}s receive \textit{bipolar-signals}, without significant corresponding \textit{bipolar-signals} received by the `surround' transistors \si{M_{Si}}s, the voltage accumulated on the capacitor \si{C_{int}} does not experience a significant discharging path through the `surround' pixels, resulting in higher voltage that pulls the output of the buffer high. The generated spike from the buffer, thus, represents the output \textit{OMS-feature-spike}, indicating an object motion detected in the `center' region with respect to the `surround' region. Timing waveforms obtained by simulation of the proposed OMS circuit based on APS pixels on GlobalFoundries 22nm node is shown in Fig. \ref{oms}(b)-(c). For illustration purposes, Fig. \ref{oms}(b) assumes all the `center' transistors $M_{Ci}$ have received a \textit{bipolar-signal} and hence are ON. The `surround' transistors $M_{Si}s$ are made ON such that 10\% of `surround' transistors are ON initially and then the number of ON `surround' transistors increase by 10\% until all the `surround' transistors are ON. The resulting voltage at the node $C_{int}$ is shown in red. As expected, the voltage on node $C_{int}$ decreases as higher percentage of `surround' transistors are switched ON. Only when sufficient `surround' transistors are ON the voltage at the node $C_{int}$ is pulled low enough to result in a low voltage at the buffer output. 

\begin{figure}[!b]
\centering
\includegraphics[width =.4\linewidth]{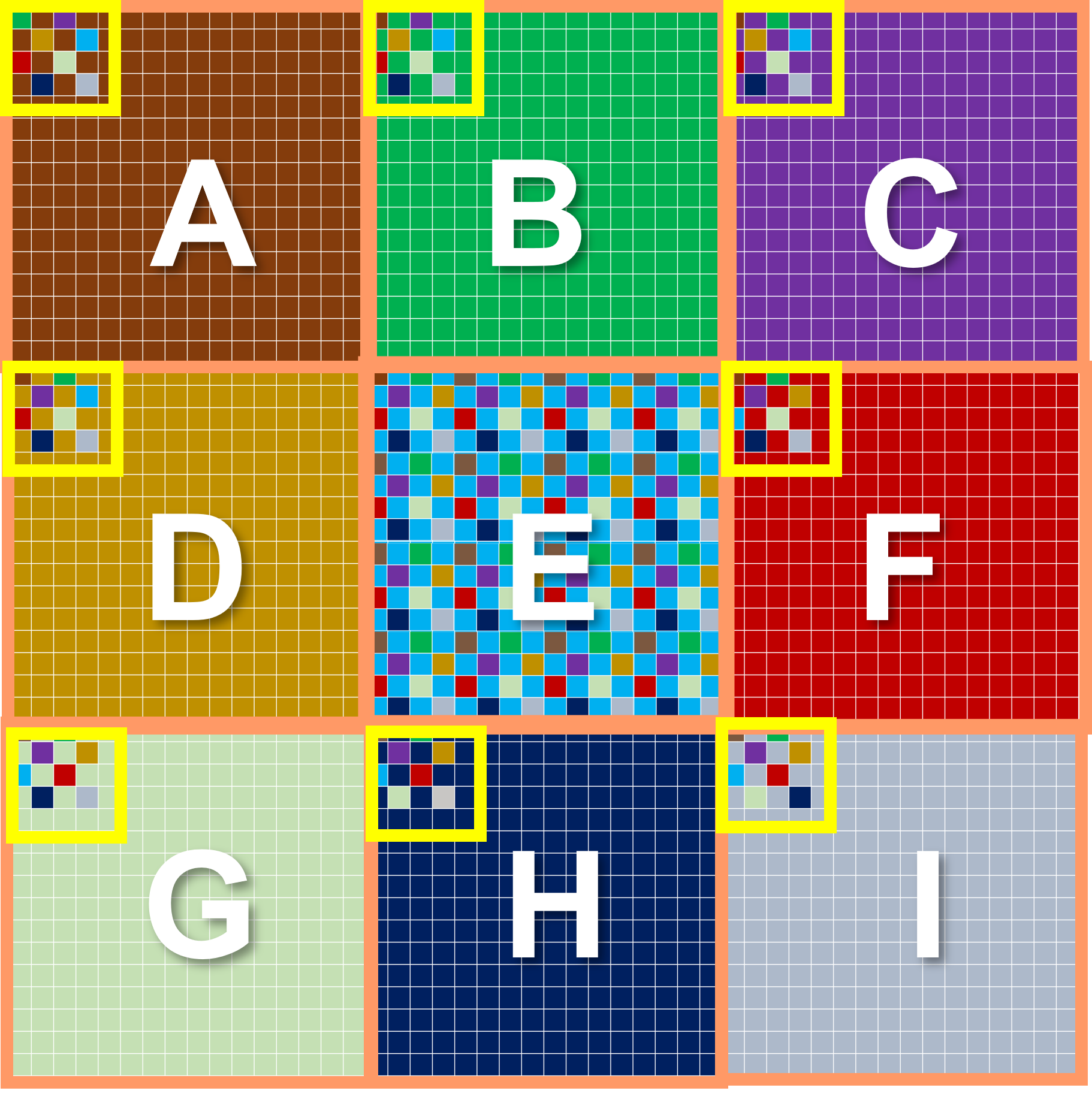}
\caption{Implementation of center-surround receptive field in a 2D array of pixels.}
\label{centersurr}
\vspace{-1mm}
\end{figure}

We would now highlight the key design aspects of the circuit proposed in Fig. \ref{oms}(a) and its connection with corresponding retinal OMS circuit of Fig. \ref{biocircuit}(a). The amacrine cells pool over larger `surround' area as compared to the 'center' area, this corresponds to higher number of `surround' transistors \si{M_{Si}} compared to the `center' transistors \si{M_{Ci}}. Pooling spikes from multiple pixels in the `surround' region is ensured in circuit of Fig. \ref{oms}(a), since all the `surround' pixels when activated drive the same capacitance \si{C_{int}}. Further, since the `surround' region is significantly larger than the `center' region, the signal generated from the `surround' region needs to be appropriately weighted by the synaptic connections to ensure proper OMS functionality. In the circuit of Fig. \ref{oms}(a), this is ensured by designing `surround' transistors \si{M_{Si}} with lower transistor widths as compared to the center transistors \si{M_{Ci}}. Finally, as shown in Fig. \ref{biocircuit}(a), the synaptic connections between amacrine cells from the `surround' region and the RGC are inhibitory in nature, while the synaptic connections between bipolar-cells in the `center' region and the RGC are excitatory in nature. We ensure such inhibitory and excitatory connections by connecting the drain of `center' pixels to \si{V_{DD}}, and the source of of `surround' pixels to ground. 

Finally, the \textit{center-surround} receptive field necessary for OMS functionality can be implemented in image sensors as shown in Fig. \ref{centersurr}. Fig. \ref{centersurr} shows a two-dimensional array of pixels. It is important to note that state-of-the-art cameras consist of millions of pixels constituting the focal plane array. We propose to divide the pixel array into multiple regions. Each individual region would act as a `center' region. For example, Fig. \ref{centersurr} shows the pixel array consisting of 9 center regions labelled $A$ through $I$. Consider a specific `center' region, say region E. The `surround' region corresponding to the `center' region E is implemented as pixels that are interleaved in the neighboring `center' regions. Specifically, the pixels corresponding to the `center' region E are represented in blue color. The `surround' pixels corresponding to the `center' region E are depicted as blue pixels embedded in the regions $A$ through $I$ except $E$. Thus, for the entire array of pixels, each `center' region will consist of majority of pixels constituting its own `center' region and fewer interleaved pixels that would correspond to the `surround' region of neighboring `center' regions. Note, the `surround' pixel interleaving is shown explicitly for all the pixels in the `center' region $E$, while it is only shown partially for the `center' region $A$ through $I$ except $E$, for visual clarity. It is worth mentioning, that the proposed method of Fig. \ref{centersurr} to mimic the \textit{center-surround} receptive field is amenable to state-of-the-art high-resolution cameras that inherently consists of numerous high-density pixels. Furthermore, the metal wires and transistors needed for routing signals between `center' and corresponding `surround' regions can be implemented using the back-end-of-line metal layers and front-end-of-line transistors from the sensor and processing die, respectively, as represented in Fig. \ref{overview}. Essentially, the backside illuminated CMOS sensor, and the heterogeneously integrated processing chip allows transistors and photodiodes to be integrated on top of sensor chip (which receives incident light) and another set of transistors that can be fabricated towards the bottom of the processing chip, with several layers of metals between them. Such a structure is naturally amenable to complex routing of signals as needed by \textit{center-surround} receptive field for  OMS functionality.

\begin{figure}[!h]
\centering
\subfloat[]{\includegraphics[width=1\linewidth]{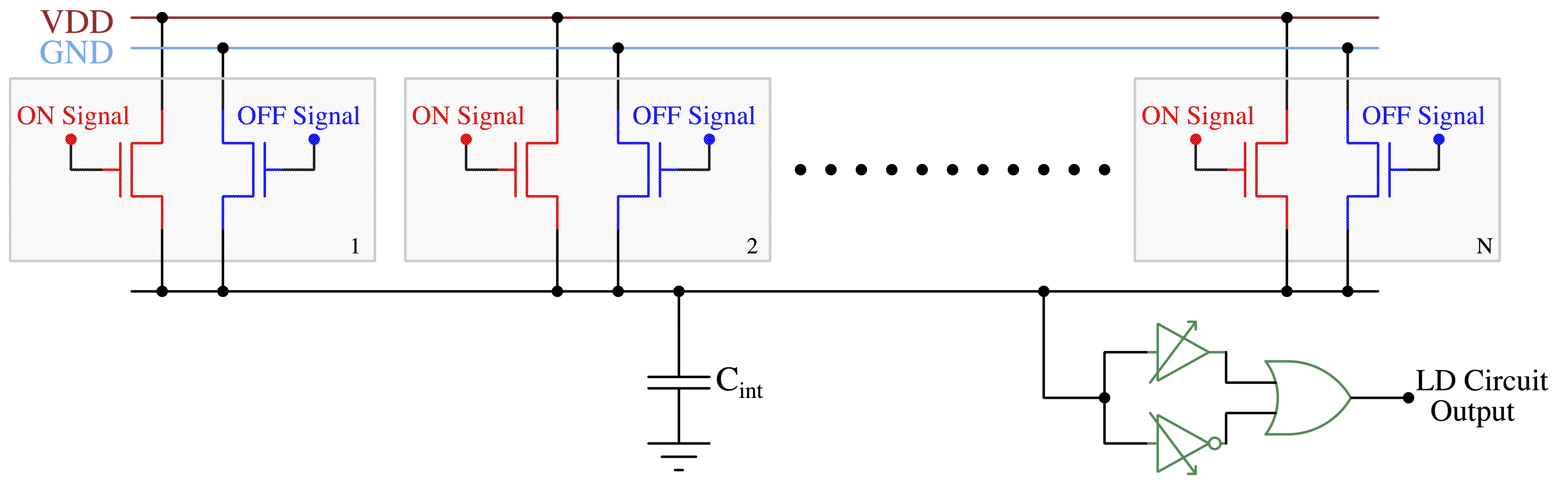}}
\newline
\centering
\subfloat[]{\includegraphics[width=0.8\linewidth]{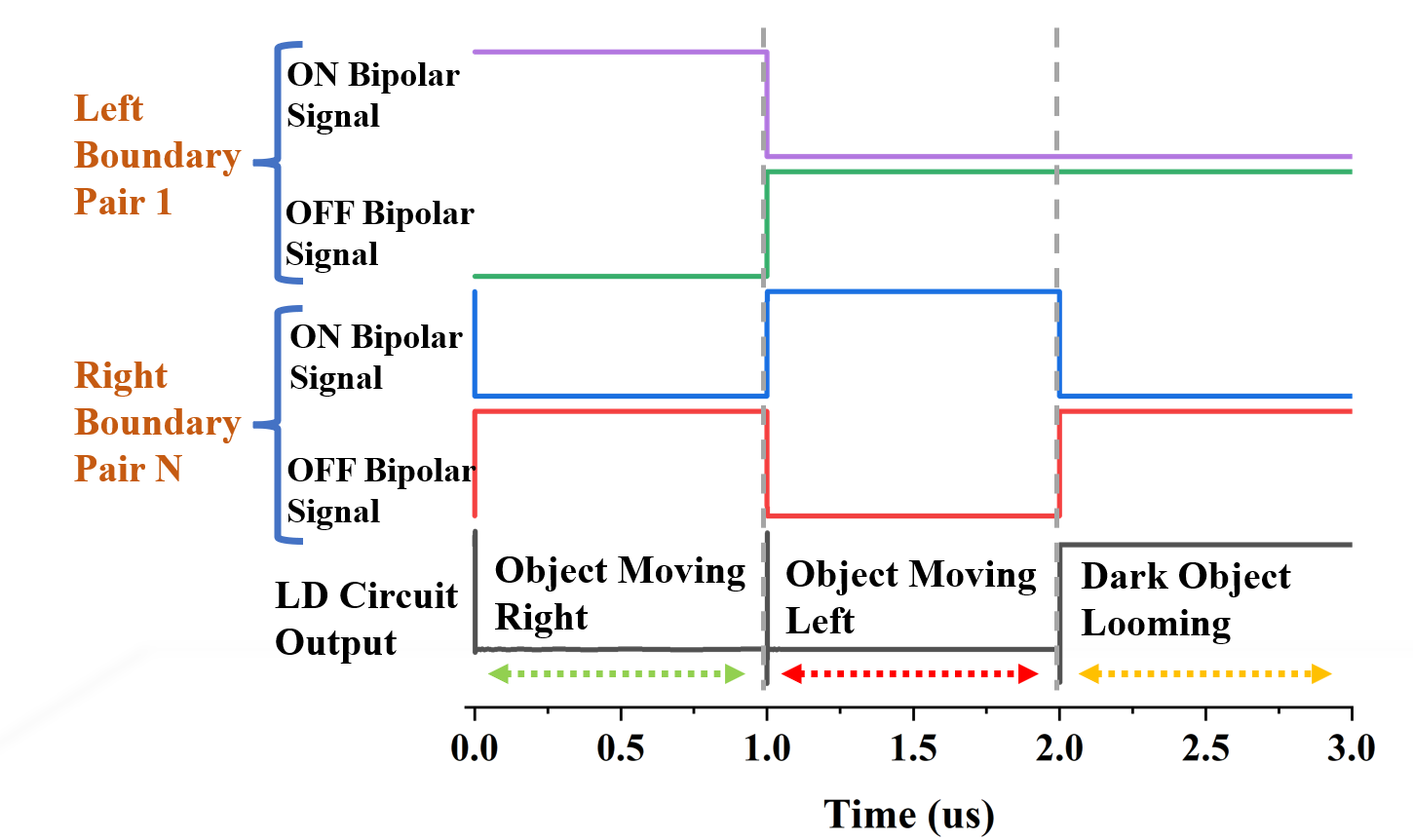}}
\caption{(a) CMOS implementation of Looming Detection Circuit diagram and (b) Timing waveform of the retinal Looming functionality showing the output voltage for three different scenarios.}
\label{lum}
\end{figure}

\section*{Embedding LD Functionality in Image Sensors }

A solid-state implementation of retinal LD circuit from Fig. \ref{biocircuit}(b) is shown in Fig. \ref{lum}(a). The figure consists of multiple pair of transistors connected to a common capacitor $C_{int}$. Each pair consists of a transistor shown in red and another in blue. The upper terminal (drain) of the red transistors are connected to $V_{DD}$, while the upper terminal (source) of transistors in blue are connected to ground. Further, the gates of the red transistors are driven by ON \textit{bipolar-signals} and the gates of the blue transistors are driven by OFF \textit{bipolar-signals}. Consider a dark object laterally moving in the receptive field. No \textit{bipolar-signals} would be generated from those pixels in the receptive field that correspond to the internal region of the dark object. This is because \textit{bipolar-signals} are only generated in response to change in light contrast. The internal region (or the body) of the dark object would continuously present low light intensity and hence would not excite any \textit{bipolar-signals}. In contrast, pixels at the boundary of the object would experience contrast change as the object moves laterally. Specifically, if the dark object is moving to the right, considering Fig. \ref{lum}(a), the pair of red and blue transistors at the left boundary of the object would experience change in light contrast. As the dark object moves to the right, the corresponding pixel pair would experience an increase in light intensity, an ON \textit{bipolar-signal} would thus be generated. The ON \textit{bipolar-signal} would activate the red transistor among the pair of transistors at the left boundary of the object. Similarly, on the right boundary of the object an OFF \textit{bipolar-signal} would be generated as the pixels at the right boundary would experience decrease in light intensity as the object moves to the right. Consequently, an OFF \textit{bipolar-signal} would be generated. The red transistor connected to the ON \textit{bipolar-signal} at the left boundary of the object would try to pull up the voltage on the capacitor $C_{int}$, while the blue transistor receiving the OFF \textit{bipolar-signal} on the right would try pulling down the voltage on the capacitor. This would result in voltage on capacitor $C_{int}$ close to $V_{DD}$/2. The logic circuit connected to the capacitor $C_{int}$ is designed to generate a low output when the voltage on $C_{int}$ is within a range of $V_{DD}$/2. The output of the logic circuit is high only when the voltage on $C_{int}$ deviates significantly from $V_{DD}$/2 (i.e. either is closer to $V_{DD}$ or closer to ground). In accordance with its behavior, the logic circuit would generate a low output in response to a voltage of $V_{DD}$/2 on node $C_{int}$ as the object moves to the right. Similar argument holds true when an object in the receptive field moves to the left, resulting in a low response from the logic circuit.

Now, consider the dark object within the receptive field is approaching (or looming). In such a case, the pair of transistors on the left and the right boundary of the object would simultaneously experience decrease in light intensity, thereby generating OFF \textit{bipolar-signals}. The blue transistors at the left and the right boundary would be activated by the OFF \textit{bipolar-signals}, while all the other transistors would remain OFF. As such, the boundary blue transistors would pull the voltage across $C_{int}$ low. In response to a low voltage on $C_{int}$ the logic circuit would generate a high output voltage (or an \textit{LD feature-spike}) indicating an approaching or looming object in the receptive field. Note, instead of a dark object the LD circuit would also generate an \textit{LD feature-spike} if a bright object is approaching in the receptive field. In this case, the red transistors at the left and right boundary of the object would be active, the node voltage on $C_{int}$ would increase closer to $V_{DD}$ and the logic circuit would respond by generating a high output. 

In accordance with the above description, Fig. \ref{lum}(b) depicts three scenarios, in the first scenario the object is moving to the right . This leads to generation of ON \textit{bipolar-signals} from the lagging edge or left boundary of the object from the pair of transistors corresponding to the receptive filed at the left boundary (Pair 1 in Fig. \ref{lum}(b)). Additionally, OFF bipolar-signals are generated from the leading edge or right boundary of the object from the pair of transistors corresponding to the receptive filed at the right boundary (Pair N in Fig. \ref{lum}(b)). The LD circuit output in this case stays low. Similar argument holds true when the object is moving to the left. However, for an approaching object (in case of Fig. \ref{lum}(b), an approaching dark object)) OFF \textit{bipolar-signals} are generated from both the left and right boundary of the object, leading to a high voltage at the output of the LD circuit.

\section*{Discussions: Future Work and Broader Impact} 

The overarching goal of IRIS sensors is to embed retinal feature extraction behavior using retina-inspired circuits within image sensors. While the current manuscript presents two key retinal functionality embedded within image sensors - Object Motion Sensitivity and Looming Detection, similar circuit-technology design techniques can be used to embed a rich class of retinal functionality including color, object orientation, object shape, and more. Some specific design considerations for IRIS sensors are as follows. IRIS sensors can be implemented based on underlying APS or DVS pixels. Specifically, for APS pixels to achieve high dynamic range a coarse grained (at pixel-array-level) or fine-grained (at individual pixel level) exposure timing control would be required \cite{zhang2020closed},\cite{sasaki2007wide}. For cameras with high density APS pixels, pixels corresponding to IRIS circuits can be scattered within typical RGB (red, green, blue) pixels, while a 3D integrated chip can house the transistors and routing metal layers for implementing corresponding IRIS circuits. This would ensure the resulting IRIS camera is capable of capturing high-resolution images, while simultaneously performing IRIS computations on the visual scene being captured by the camera. Further, the photodiodes associated with IRIS sensors could span a wide range of wavelengths including visible light \cite{xu2005backside}, near infra-red light \cite{kaufmann2011near}, and infra-red \cite{peizerat201288db}.

A key technology-enabler for IRIS sensors is advances in 3D integration of semiconductor chips. 3D integration allows integration of routing metal layers and transistor based circuits required for implementing spatio-temporal computations directly above (or under) the pixel array, similar to the biological retinal circuit. Such 3D integrated IRIS sensors can use various 3D packaging technologies like metal-to-metal fusion bonding \cite{raymundo2013exploring}, TSVs \cite{coudrain20133d}, etc. Further, heterogeneous sensors operating at different wavelengths can be co-integrated to extract retina-like feature vectors over different spectrum of light. Additionally, emerging non-volatile technologies like Resistive Random Access Memories (RRAMs) \cite{zahoor2020resistive}, Magnetic Random Access Memories (MRAMs) \cite{apalkov2013spin}, Phase Change Memories (PCM) \cite{lacaita2008phase}, Ferro-electric Field Effect Transistors (Fe-FET) \cite{lue2002device}, etc. can be used for IRIS circuits, for example, to implement programmable weights for `center' and `surround' regions. 

Lastly, IRIS sensors could have significant impact on computer vision in general. Today's computer vision exclusively relies on light intensity-based (APS) or light change-detection-based (DVS) pixels data collected through state-of-the-art CMOS image sensors. However, in almost all cases, appropriate context for the pixels is missing (or are extremely vague) with respect to the `real-world events’ being captured by the sensor. Thus, the onus of processing is rather put on intelligent machine learning algorithms to pre-process, extract appropriate context, and make intelligent decisions based on pixel data. Unfortunately, such a vision pipeline leads to 1) complex machine learning algorithms designed to cater to image/video data without appropriate context 2) increases the time to decision, associated with machine learning algorithms requiring to process millions of pixels per frame 3) energy-hungry and slow access to pixel data being captured and generated by the CMOS image sensor. IRIS sensors could usher in new frontiers in vision-based decision making by generating highly specific motion and shape-based features, providing valuable context to pixels captured by the camera. The underlying algorithms processing data generated from IRIS sensors could be based on traditional deep learning models or on emerging set of spiking neural networks that could process \textit{feature-spikes} generated from IRIS sensors. Finally, since IRIS cameras can use APS pixels, in general, they can generate both feature-spikes and light intensity map as and when required by the computer vision algorithms.

\section*{Materials and Methods} Hardware circuit simulations were performed using process design kit (PDK) from Globalfoundries for fully-depleted Silicon-on-insulator (FD-SOI) technology at 22nm technology node \cite{globalfoundries_2022}. The 22nm FD-SOI node is well suitable for the kind of mixed signal circuits used in this work \cite{globalfoundries_2015}.  The simulations were run on industry standard EDA (Electronic Design Automation) tools from Cadence \cite{cadence}.

Algorithmic implementations of the OMS and LD circuits (Fig.  \ref{ExampleFrames}) were performed in MATLAB. Code is available at \textit{https://github.com/SchwartzNU/ViBES}.

\section*{Acknowledgments}
The work was supported in part by Keston Foundation Exploratory Research Award and Center for Undergraduate Research at Viterbi (CURVE), University of Southern California, Los Angeles, USA.





\bibliography{scibib} 
\bibliographystyle{science}

\end{document}